\documentclass{PoS}

\usepackage{latexsym} 	
\usepackage{amsmath}  	
\usepackage{amsbsy}
\usepackage{epsfig}     
\usepackage{cite}

\usepackage{epsfig}
\usepackage{graphicx}
\usepackage{dcolumn}
\usepackage{bm}

\newcommand \etc {{\sl etc.\ }}

\newcommand \ie {{\sl i.e.\ }}

\newcommand{\be}{\begin{equation}}
\newcommand{\ee}{\end{equation}}
\newcommand{\bea}{\begin{eqnarray}}
\newcommand{\eea}{\end{eqnarray}}

\title{Chiral phase transition of $N_f$=2+1 QCD with the HISQ action }

\ShortTitle{Chiral phase transition of $N_f$=2+1 QCD with the HISQ action}

\author{\speaker{H.-T. Ding}$^{1,2}$\footnote{Current address: Institute of Particle Physics, Central China Normal University, Wuhan, 430079, China}~, 
        A. Bazavov$^3$, 
        F. Karsch$^{1,4}$, 
        Y. Maezawa$^{4}$, 
        Swagato Mukherjee$^1$ and P. Petreczky$^1$\\
        \llap{$^1$}
Physics Department, Brookhaven National Laboratory, Upton, NY 11973, USA \\
        \llap{$^2$}
        Physics Department, Columbia University, New York, NY 10027, USA \\
        \llap{$^3$}
        Department of Physics and Astronomy, University of Iowa, Iowa City, IA 52240, USA\\
        \llap{$^4$}
     Faculty of Physics, University of Bielefeld, 
        D-33501 Bielefeld, Germany\\
        E-mail: 
        \email{HengTong.Ding@mail.ccnu.edu.cn}
        }

\abstract{
We present studies of universal properties of the chiral phase transition in $N_f$=2+1 QCD based on the simulations using Highly Improved Staggered fermions on lattices with temporal extent $N_\tau$=6. We analyze the quark mass and volume dependence of the chiral condensates and chiral susceptibilities in QCD with two degenerate light quarks and a strange quark. The strange quark mass is chosen to be fixed to its physical value ($m^{phy}_s$) and five values of light quark masses ($m_l$) that are varied in the interval 1/20$\gtrsim m_l/m^{phy}_s \gtrsim$1/80. Here various quark masses correspond to pseudo Goldstone pion masses ranging from about 160 MeV to about 80 MeV. The O(N) scaling of chiral observables and the influence of universal scaling on physical observables in the region of physical quark mass values are also discussed. 
}

\FullConference{31st International Symposium on Lattice Field Theory - LATTICE 2013\\
		July 29 - August 3, 2013\\
		Mainz, Germany}

\begin{document}

\maketitle

\section{Introduction}

\begin{figure}[htp]
\begin{center}
\includegraphics[width=0.35\textwidth]{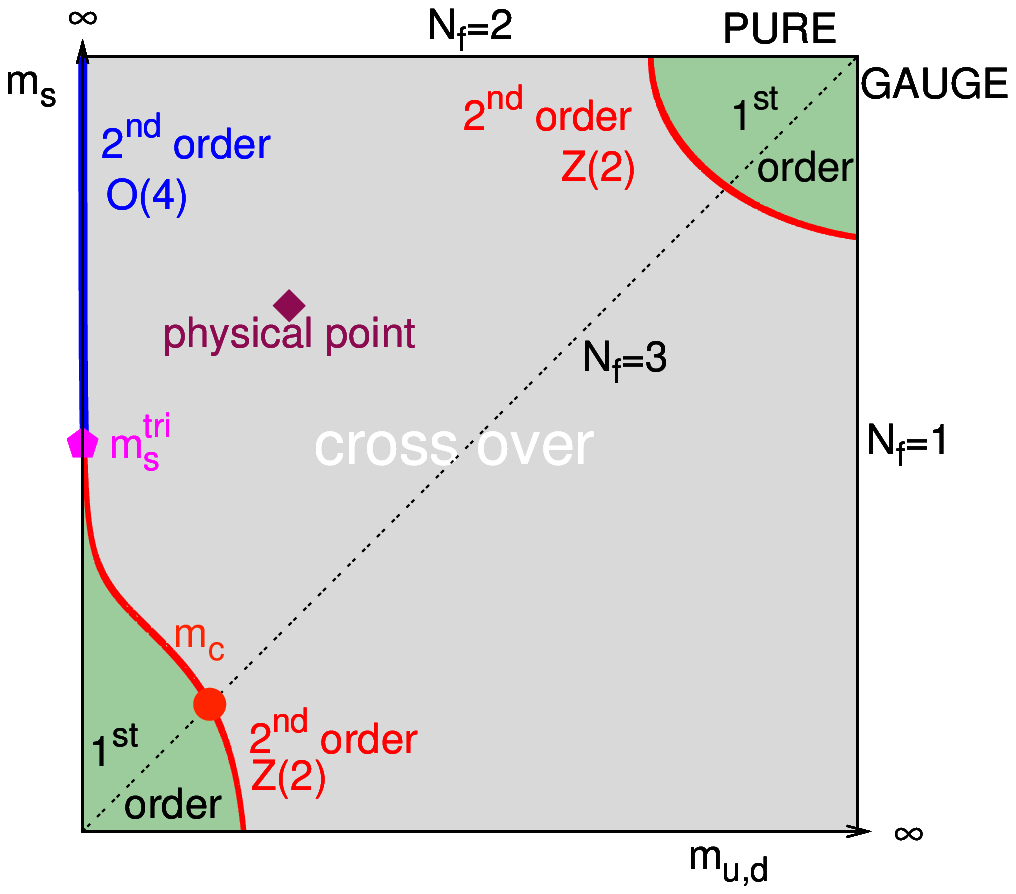}~~~
\includegraphics[width=0.45\textwidth,height=0.28\textwidth]{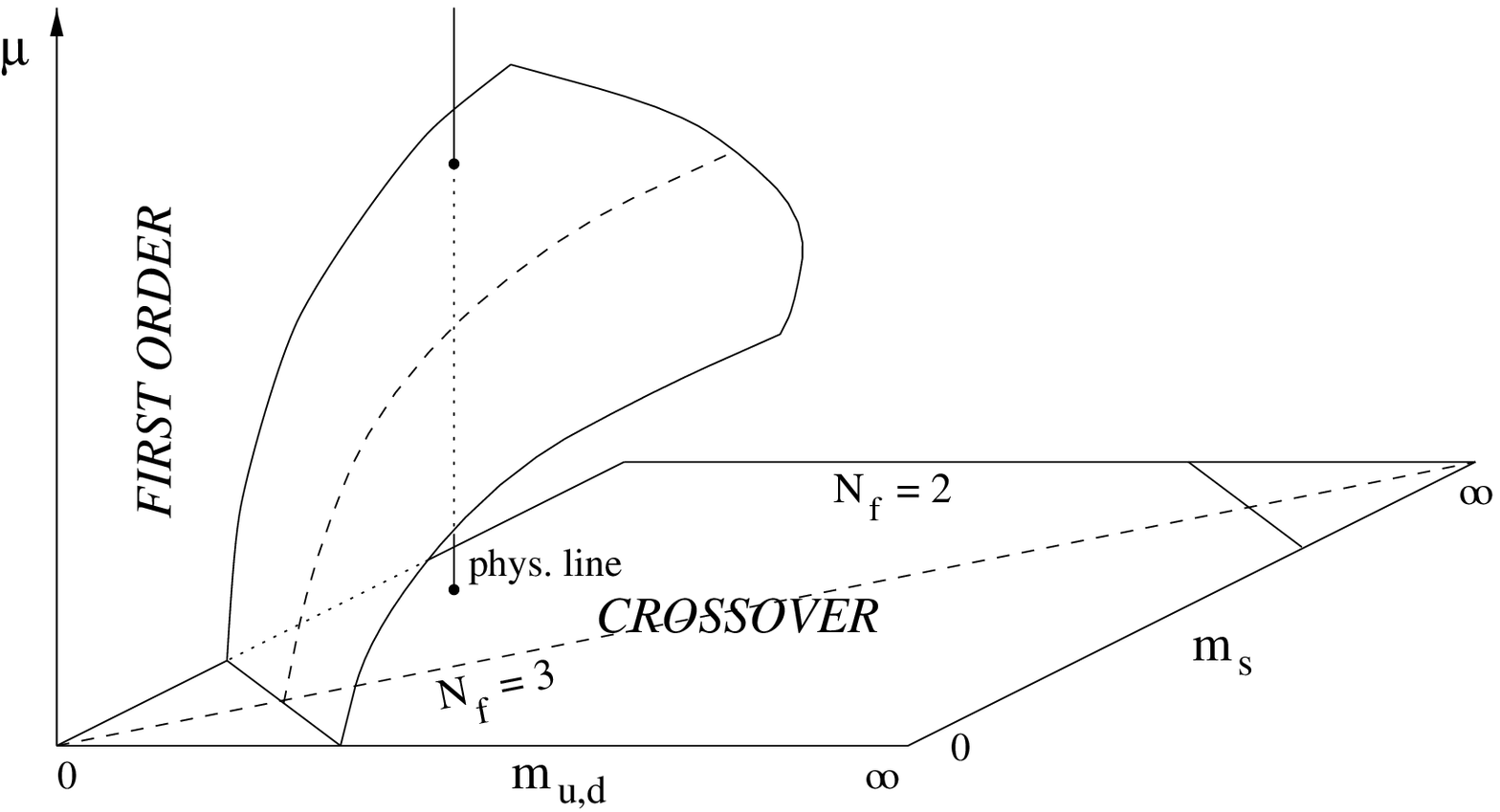}
\end{center}
\caption{Left: schematic QCD phase transition behavior for different choices of quark 
masses ($m_{u,d}$, $m_{s}$) at zero chemical potential. Right: The critical 
surface swept by the chiral critical line at finite chemical potential. A QCD chiral critical point may exist if the surface bends towards to the physical point. The right 
plot is taken from Ref.~\cite{Karsch04}.}
\label{fig:sketch}
\end{figure}

The exploration of the QCD phase diagram is one of the basic goals of lattice
QCD calculations at non-zero temperature. It had been noted by Pisarski and 
Wilczek that the order of phase transitions in QCD may depend on the number of
light quark degrees of freedom and that qualitative features of the transition
may also change with the quark mass \cite{Pisarski:1983ms}. 
The basic features of these considerations have been verified in lattice
QCD calculations and generally are summarized in the so-called Columbia plot~\cite{Brown:1990ev},
which shows the different regions of first order, second order and crossover
transitions as a function of the two degenerate light and the strange quark mass
(see Fig.~\ref{fig:sketch}~(left)~\cite{Ding:2013nha}).

It now seems to be well established that the transition in QCD with its physical mass
spectrum is just a crossover~\cite{Wuppertal06,HotQCDTc}.  In the chiral limit of $N_f=2+1$ QCD\footnote{The
chiral limit in this case is taken for fixed physical values of the strange quark
masses.} the chiral transition of QCD could be either first or second order depending on the 
value of strange quark at the tri-critical point ($m_s^{tri}$) and the breaking strength of the $U(1)_A$ symmetry~\cite{Pisarski:1983ms}.
There is some evidence
that the QCD phase transition in the chiral limit of (2+1)-flavor QCD is second order and belongs to the universality class of three dimensional
$O(N)$ spin models~\cite{ejiri-1}\footnote{In the continuum limit, taken before the
chiral limit is taken, the universality class will be $O(4)$. For studies with
staggered fermions at finite values of the lattice cut-off the appropriate symmetry
group is $O(2)$.}.  However, the currently existing $O(N)$ scaling studies have been
performed on rather coarse lattices and with staggered fermion actions that are no
longer state-of-the-art, i.e. lead to large taste violations. The lattice studies using chiral fermions, which reproduce correct chiral and axial symmetries of QCD,  e.g. Domain Wall Fermions,
are currently too expensive to go down to lower than physical quark masses to have any detailed studies on the scaling behavior of 
the chiral phase transition~\cite{Buchoff:2013nra}.
It thus is not
surprising that the order of the QCD phase transition in the chiral limit is still
under debate and arguments in favor of a first order transition have been put forward
\cite{Aoki12}.

The chiral transition in QCD at vanishing quark chemical potentials is one of the
corner stones for our understanding of non-perturbative effects in strongly
interacting matter. A large experimental program is put forward to understand and
find evidence for its existence in various observables. In order to understand how
the chiral phase transition at vanishing quark masses influences properties of QCD with its physical mass spectrum,
it is crucial to get control over the continuum and chiral limit and disentangle
universal, critical behavior in QCD thermodynamics from ordinary regular
contributions to the QCD partition function. Getting control over the structure of
the phase diagram at vanishing chemical potential will also influence our thinking
about its extension to non-zero chemical potential. For instance, it has long been
argued that the second order boundary line of the region of first order transition
in the small quark mass region of 3-flavor QCD may play a role in the analysis of a
possible critical point in QCD at non-zero density (see Fig.~\ref{fig:sketch} (right)). We now
learned that this region is tiny and detached from the physical mass
regime~\cite{Ding:2013nha,Endrodi:2007gc}. The behavior of QCD in the chiral limit of
(2+1)-flavor QCD thus becomes more relevant for our understanding of the critical
behavior at zero and non-zero baryon chemical potential. An important issue is
to establish the dependence of the QCD transition temperature on the baryon chemical
potential and analyze its relation to the experimentally determined freeze-out curve. In order to have a well defined definition of the transition
line it is essential to use a prescription that relates to the true second order
phase transition line in the chiral limit. This too requires the analysis of the
quark mass dependence of various observables, e.g. mixed susceptibilities, that can
be used to locate the pseudo-critical temperature in QCD with its physical quark mass
spectrum. 

In this proceedings we report on the current state of the art investigation on the chiral phase transition of $N_f$=2+1 QCD using the Highly Improved Staggered Quarks.

\section{Lattice parameters}

The Highly Improved
Staggered Quark (HISQ) action \cite{HISQ} action has already been used for the study of QCD thermodynamics and it has been found that it significantly
reduces the lattice artifacts caused by the taste-symmetry specially in the
transition region \cite{HotQCDTc}.  The use of the HISQ action with these lattice
spacings is expected to reduce the taste-symmetry breaking by more than an order of
magnitude \cite{HotQCDTc} compared to the existing state-of-the-art computations of
chiral observables shown in Ref. \cite{ejiri-1}. 

\begin{table}[t]
\small
\begin{center}
\begin{tabular}{|c|c|c|c|c|}
\hline
lattice dim. &   $m_l/m^{phy}_s$     & $m_{\pi}$ [MeV]     &\# $\beta$ values & no. of traj. \\
\hline
$24^3\times$ 6           & 1/20          & 160                       &  17                    & 10000             \\
$12^3\times$ 6          & 1/27        & 140                 &  11                    & 10000            \\
$16^3\times$ 6          & 1/27        & 140                    &  11                   & 10000            \\
$20^3\times$ 6          & 1/27        & 140                    &  11                    & 10000             \\
$24^3\times$ 6          & 1/27        & 140                    &  14                    & 10000             \\
$32^3\times$ 6          & 1/27        & 140                    &  11                    & 10000           \\
$32^3\times$ 6           & 1/40      & 110                      &  16                   &  8000           \\
$40^3\times$ 6          & 1/60        &  90                       &   11                 & 7000           \\
$32^3\times$ 6           & 1/80    & 80                       &   5                     & 3000           \\
$48^3\times$ 6           & 1/80    & 80                       &   5                     & 9000           \\
\hline
\end{tabular}
\end{center}
\caption{
Parameters of the numerical simulations.
}
\label{tab:parameter}
\end{table}

Given the importance of improved chiral symmetry, i.e. reduced violations of the
taste-symmetry, we performed our simulations of $N_f=2+1$ QCD with lattice
spacings corresponding to temporal lattice size $N_\tau=6$ using the HISQ action. 
In the current simulation the strange quark mass is chosen to be fixed to its physical value ($m^{phy}_s$) and five values of light quark masses ($m_l$) that are varied in the interval 1/20$\gtrsim m_l/m^{phy}_s \gtrsim$1/80. Here various quark masses correspond to the lightest pseudo Goldstone pion masses of about 160, 140, 110, 90 and 80 MeV. 
To study the volume dependences of various chiral observables
we performed simulations at the physical quark mass, i.e. $m_l/m^{phy}_s$=27, with spatial lattice size 
of $N_s=$12, 16, 20, 24 and 32. We too performed simulations at two different volumes, i.e. $N_s=48$ and 32 at our lowest quark mass corresponding to lightest Goldstone pion mass of about 80 MeV.
To ensure $m_\pi L \gtrsim 4$ we performed simulations with $N_s$=24, 32 and 40 at light quark mass $m_l$=$m^{phy}_s/20$, $m^{phy}_s/40$ and $m^{phy}_s/60$, respectively.
The parameters of simulations reported in current study are summarized in Table~\ref{tab:parameter}.

\section{Universality class near critical lines}

Close to the chiral limit the chiral order parameter ($M$) and its
susceptibility ($\chi_M$) can be described by the universal properties of the chiral
transition \cite{ejiri-1}
\be
M(t,h)= h^{1/\delta} f_G(z) 
\;, \qquad\mathrm{and}\qquad
\chi_M(t,h) = \frac{\partial M}{\partial H}=\frac{1}{h_0} \,h^{1/\delta -1}\,
f_\chi(z)
\;.
\label{eq:MEOS}
\ee
Here $z=th^{-1/\beta\delta}$ is the scaling variable, $t=\frac{1}{t_0} \,\frac{T-T_c}{T_c}$ and $h= \frac{1}{h_0}\frac{m_l}{m_s}$
are the rescaled temperature and the rescaled quark mass, respectively. 
Scaling variables $t$ and $h$ measure the distance that the system is away from the criticality.  The critical exponents $\beta$, $\delta$ and the scaling functions $f_G(z)$ and 
$f_\chi(z)$ uniquely characterize the universality class of the chiral
phase transition in 2-flavor QCD, which is believed to be equivalent to that of the 3-d O(4) spin
model. However, the parameters $t_0$, $h_0$ are non-universal and specific for a
theory, \ie they depend on the action, lattice spacing, value of the bare strange
quark mass \etc These non-universal parameters can be
determined by studying the scaling behaviors of the chiral order parameter and the
chiral susceptibility described in Eq.~(\ref{eq:MEOS}).

In the low temperature limit, which corresponds to large negative values of the scaling variable $z$, the scaling function $f_G(z)$, $M$ and $\chi_M$ behavior in the following way~\cite{Engels:2001bq}
\bea
f_G(z)\simeq f_G^{-\infty}(z) &=& (-z)^\beta \left ( 1+ c_2\, \beta (-z)^{-\beta\delta/2} \right), \\
M  \simeq  h^{1/\delta} f_G^{-\infty}(z) &=&  (-t)^\beta  \left (1+c_2\,\beta (-z)^{-\beta\delta/2} \sqrt{h} \right), ~~~~~ \chi_M \sim  h^{-1/2}.
\label{eq:On_lowT}
\eea
Thus the contribution of Goldstone modes to the order parameter is already enclosed in the scaling function.
In the high temperature limit, which corresponds to the large positive value of $z$, the scaling function $f_G(z)$, order parameter $M$ and its susceptibility
behavior as follows~\cite{Engels:2003nq}
\bea
f_G(z)  \sim R_\chi \,\, z^{-\beta(\delta -1)}, ~~~~M\sim R_\chi\,\, t^{-\beta(\delta -1)}h, ~~~~\chi_M \sim  R_\chi\,\, t^{-\beta(\delta -1)}.
   \label{eq:On_highT}
\eea
where $R_\chi$ is a universal coefficient and independent of $h$. Note that in this case $\chi_M$ is independent on $h$.

\section{Results}

\begin{figure}[htp]
\begin{center}
\includegraphics[width=0.45\textwidth]{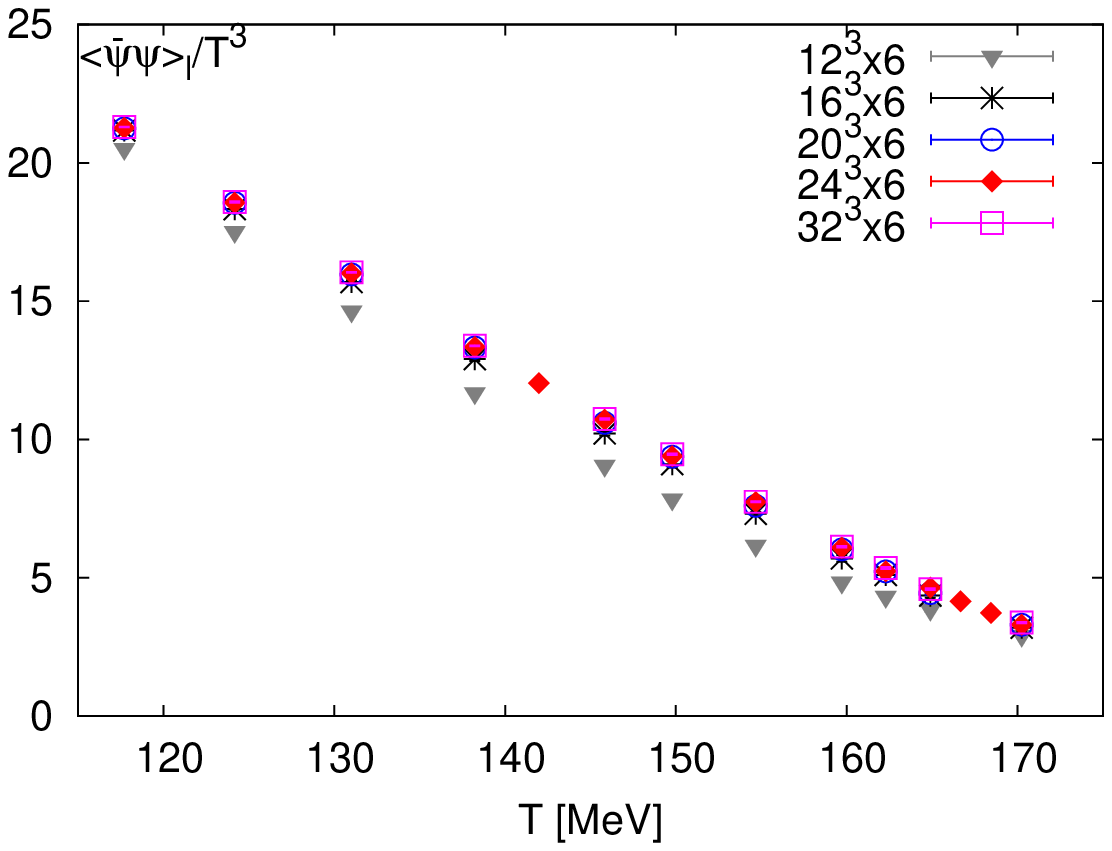}~~~
\includegraphics[width=0.45\textwidth]{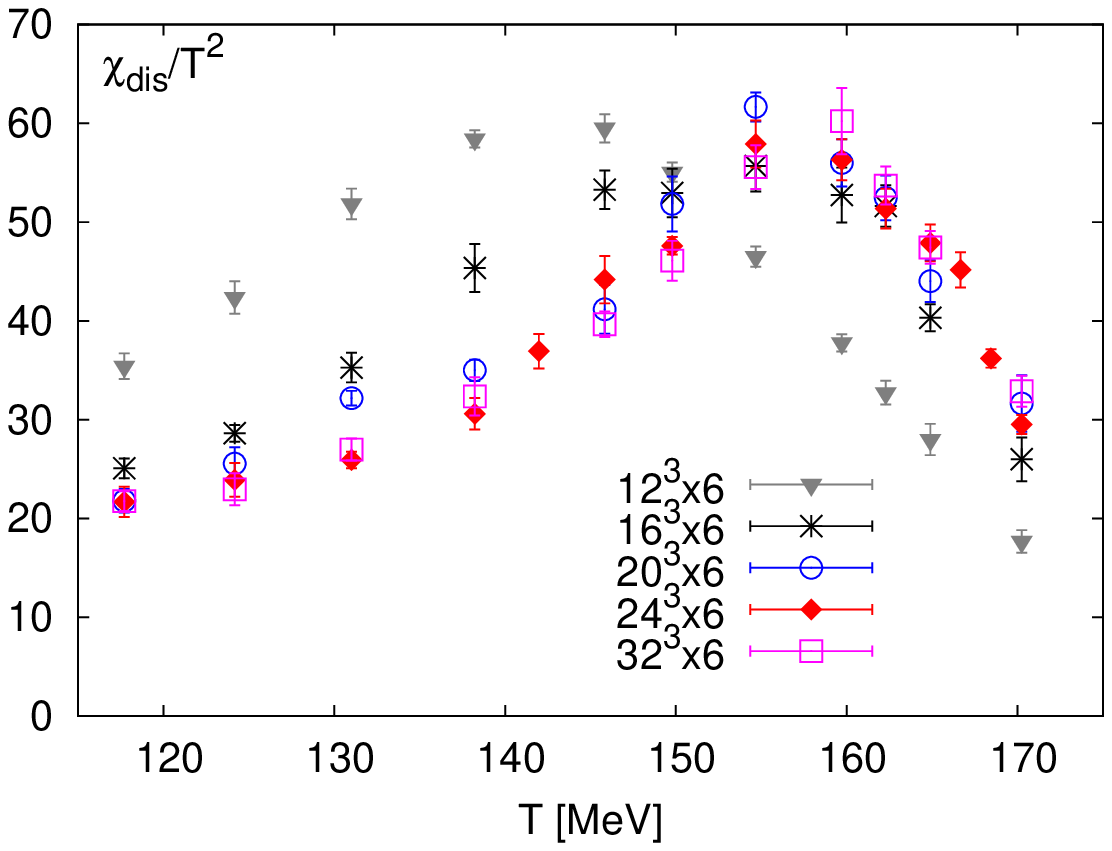}~
\end{center}
\caption{Volume dependences of light quark chiral condensates (left) and disconnected chiral susceptibilities (right) at the physical quark mass.}
\label{fig:Vdep_phys}
\end{figure}

In Fig.~\ref{fig:Vdep_phys} we show volume dependences of light quark chiral condensates $\langle\bar{\psi}\psi\rangle_l$ and disconnected chiral susceptibilities $\chi_{disc}$ at the physical quark mass in the
left and right plot, respectively. The chiral condensate have minor volume dependences when $N_s\geq20$. Together with the volume dependence of disconnected chiral susceptibilities
it is clear that for $N_s\geq 24$ at the physical quark mass the system reaches the thermodynamic limit.

\begin{figure}[htp]
\begin{center}
\includegraphics[width=0.45\textwidth]{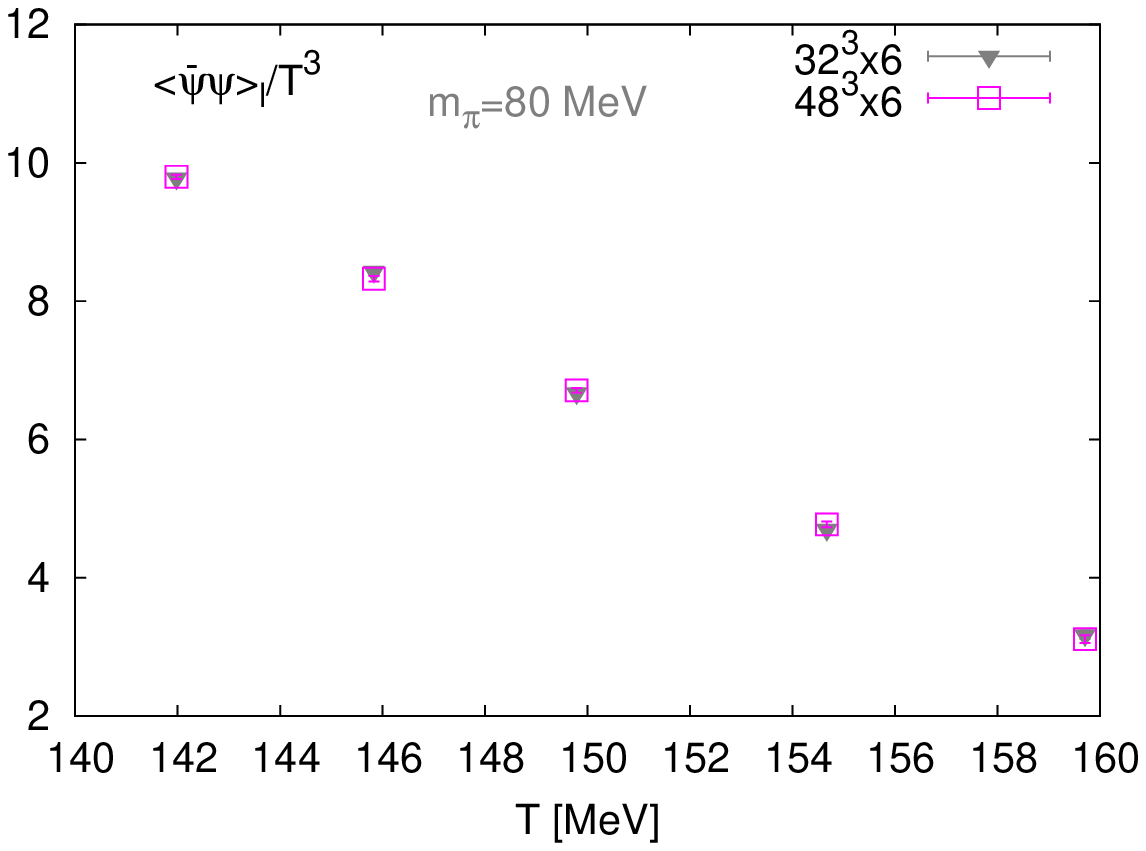}~~~
\includegraphics[width=0.45\textwidth]{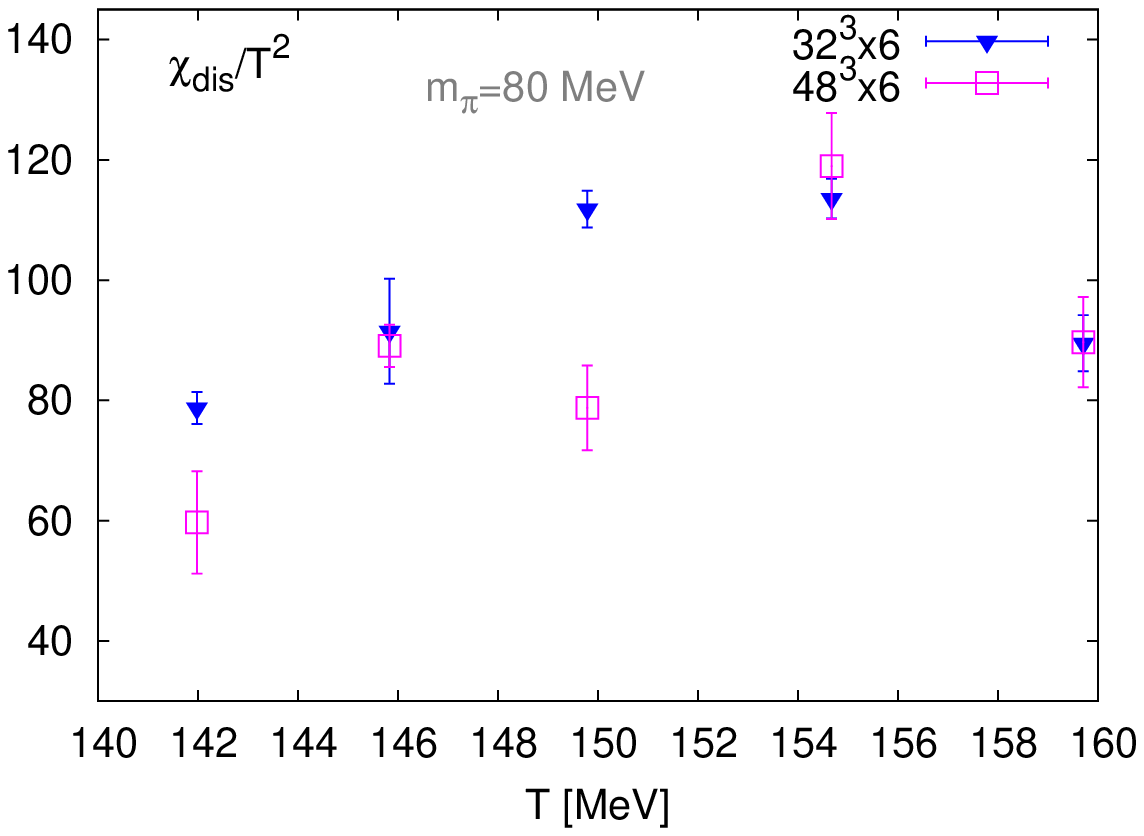}~
\end{center}
\caption{Volume dependences of light quark chiral condensates (left) and disconnected chiral susceptibilities (right) for the light quark mass $m_l=m_s^{phy}/80$ corresponding
to $m_\pi\simeq80~$MeV.}
\label{fig:Vdep_80MeV}
\end{figure}

The volume dependences of chiral condensates and disconnected chiral susceptibilities at our lowest quark mass corresponding to $m_\pi\simeq 80 $MeV are shown in Fig.~\ref{fig:Vdep_80MeV}.
No evidence of a first order phase transition of the QCD medium at current quark mass is observed. 

\begin{figure}[htp]
\begin{center}
\includegraphics[width=0.33\textwidth]{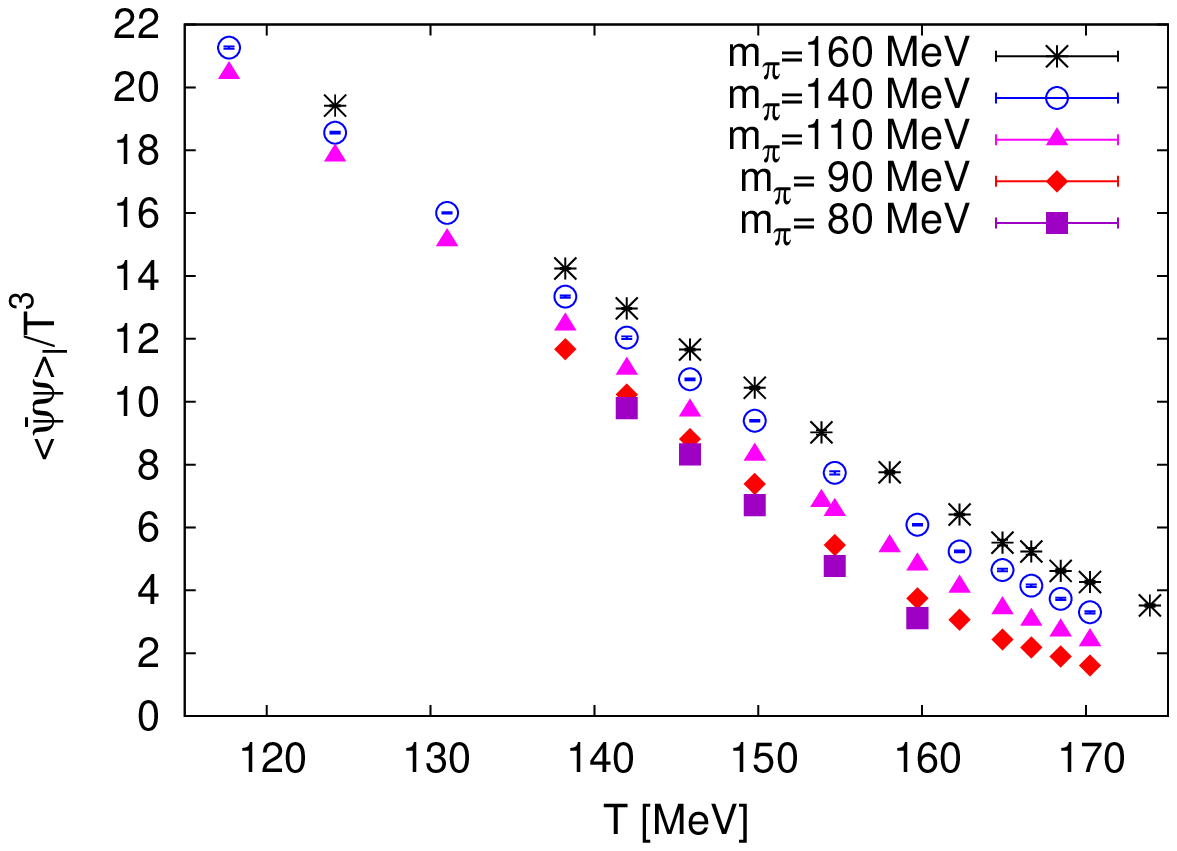}~~~
\includegraphics[width=0.33\textwidth]{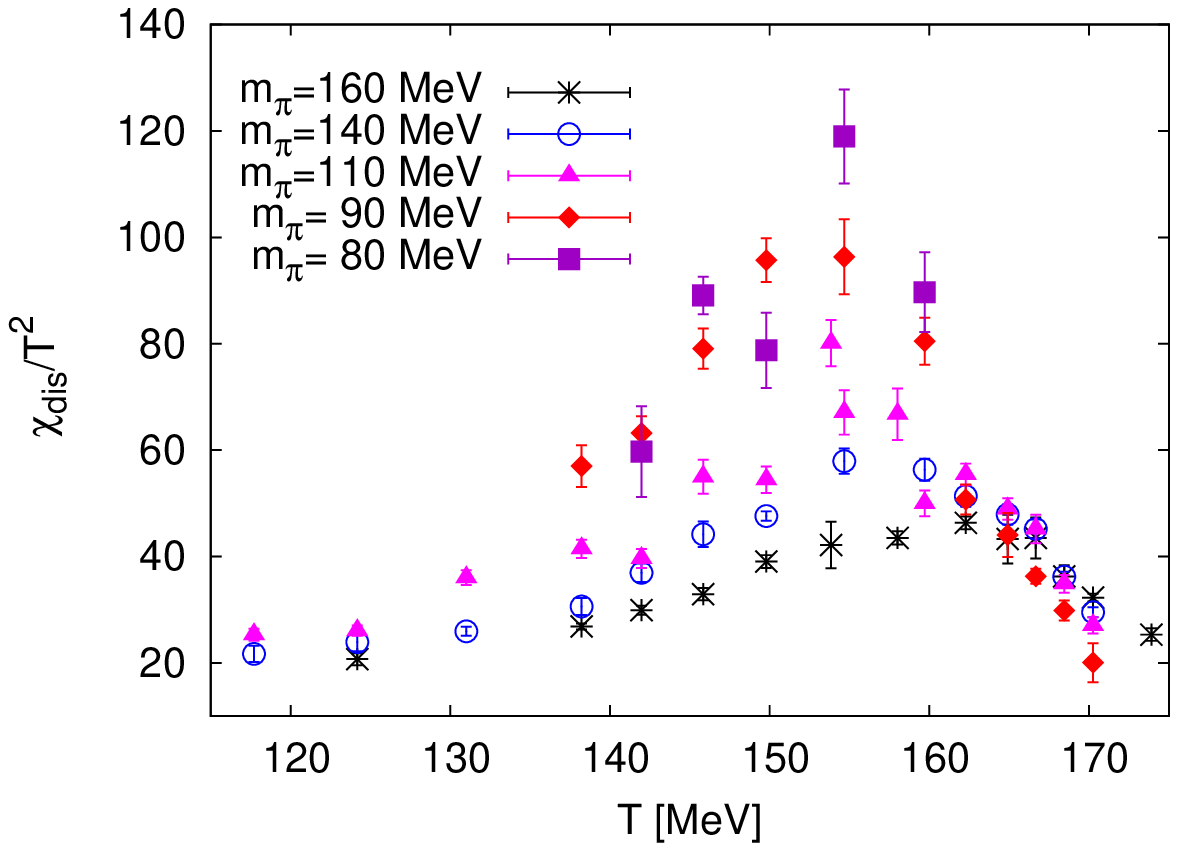}~
\includegraphics[width=0.33\textwidth]{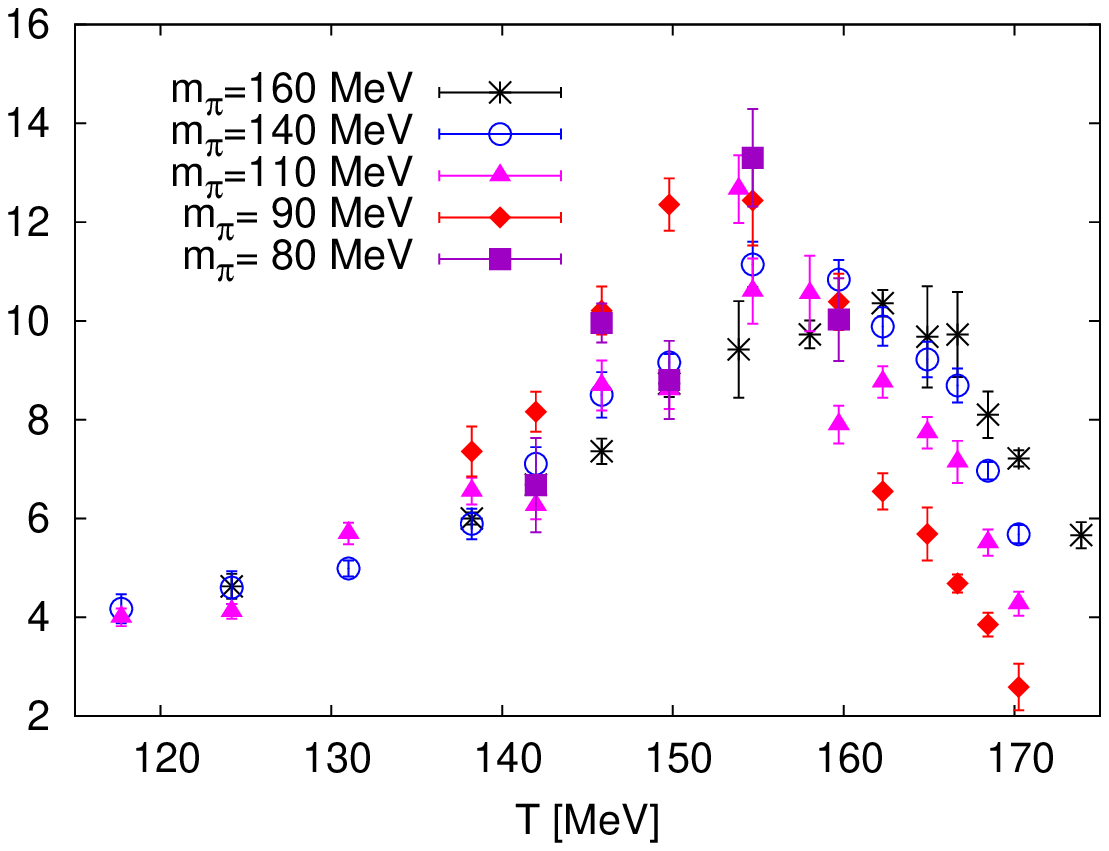}
\end{center}
\caption{Quark mass dependences of light quark chiral condensates (left), disconnected chiral susceptibilities (middle) and rescaled disconnected chiral susceptibilities by a factor of $(m_l/m^{phy}_s)^{1/2}$(right).}
\label{fig:Mdep}
\end{figure}

The mass dependences of chiral condensates and disconnected chiral susceptibilities
are shown in Fig.~\ref{fig:Mdep}. Seen from the left plot no discontinuity of chiral condensates in temperature is present in the currently investigated quark mass window.
Observed from the plot in the middle it seems that in the high temperature region disconnected chiral condensates are independent of quark mass. It may indicate the O(N) scaling
of disconnected chiral susceptibilities according to Eq.~(\ref{eq:On_highT}) although more statistics especially of the dataset for low quark mass are needed to draw firm conclusion.
In the right plot we show the disconnected chiral susceptibility multiplied by a factor of $(m_l/m^{phy}_s)^{1/2}$. In the low temperature region one can see that 
the rescaled disconnected chiral susceptibility is almost independent of quark mass. This arises from the contribution of Goldstone modes which
is encoded in Eq.~(\ref{eq:On_lowT}).

\begin{figure}[htp]
\begin{center}
\includegraphics[width=0.45\textwidth]{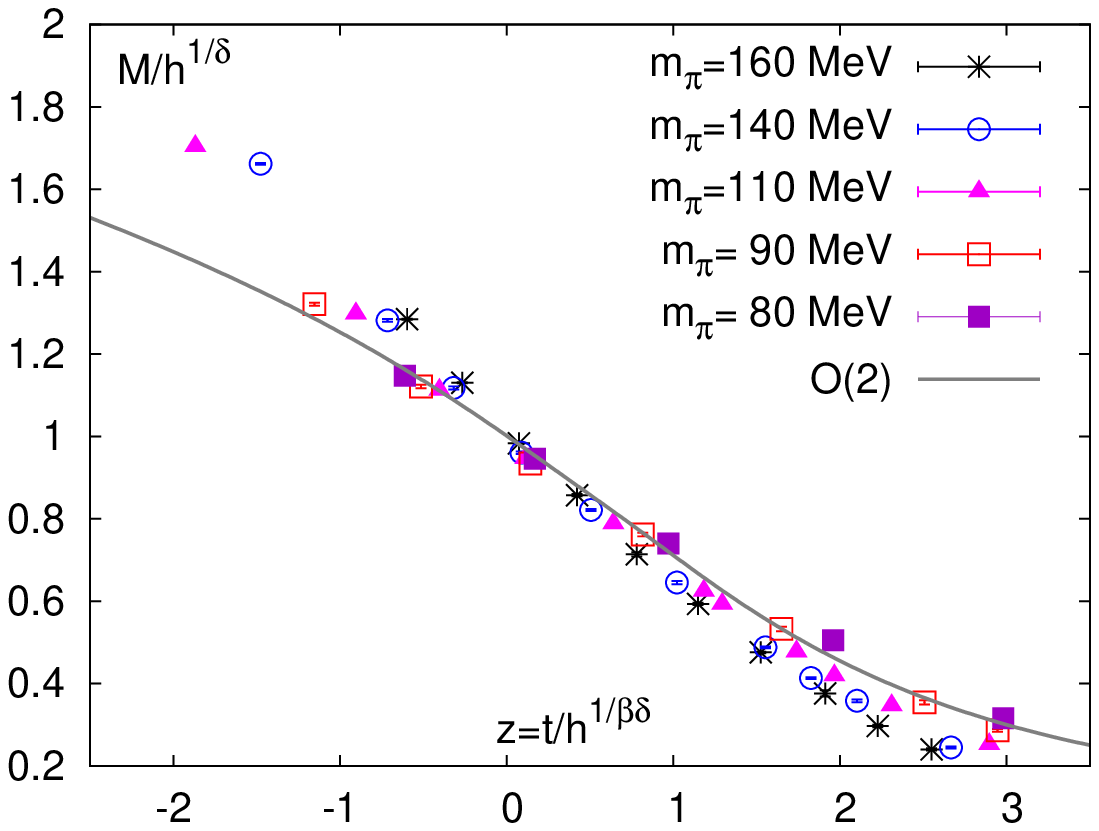}~
\includegraphics[width=0.45\textwidth]{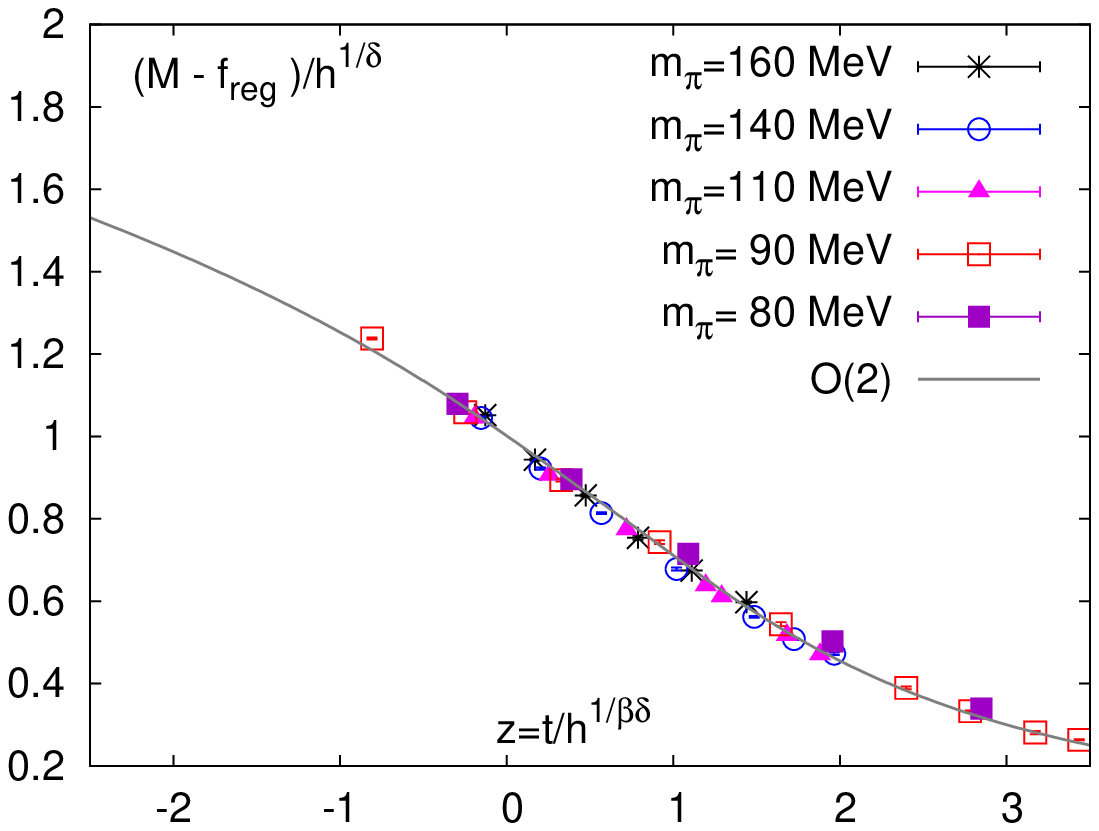}
\end{center}
\caption{Left: O(2) scaling analysis for all quark masses using
non-universal parameters $t_0$, $h_0$ and $T_c$ obtained from the scaling fits to the order parameter $M=m_s\langle \bar\psi\psi\rangle_l N_\tau^4$ for our lightest two pion masses. Right: O(2) scaling analysis with regular terms parameterized as in Eq.~(4.1).}
\label{fig:MEOS}
\end{figure}

To check the scaling window of QCD system using the HISQ action we assume our lowest two quark mass are in the scaling regime.
We fit the order parameter M defined as $m_s\langle\bar{\psi}\psi\rangle_l N_\tau^4$ at two lowest quark masses using Eq.~(\ref{eq:MEOS}) with O(2) scaling function\footnote{We also performed a fit using Z(2) scaling function with an additional parameter
$m_c$ and the fit favors a vanishing value of $m_c$.} and obtained the non-universal critical exponents
$t_0$, $h_0$ and $T_c$. We then plot all the data sets using these non-universal critical exponents in the framework of $M/h^{1/\delta}$ as a function
of $z$ as shown in the left plot of Fig.~\ref{fig:MEOS}. We can clearly see that scaling window in the case of HISQ action shrinks compared to the
p4fat3 action on $N_\tau=4$ lattices shown in Ref.~\cite{ejiri-1}. The scaling violation seen in the left plot may arise from  the regular part of the QCD partition function. 
Here we fit the scaling violation via the following ansatz
\be
M = h^{1/\delta} f_G(z) + f_{reg}, ~~~~f_{reg}= \left(a_0 + a_1 \frac{T-T_c}{T_c} \right) \frac{m_l}{m_s},
\label{eq:reg}
\ee
and the fit is shown in the right plot of Fig.~\ref{fig:MEOS}. The above ansatz with only linear behavior of quark mass in the regular term seems to be sufficient 
to describe the data and this allows a description of the scaling violation at nonzero values of quark masses via the study of the scaling behavior of the chiral phase transition.

\section{Summary}
In order to study the scaling behavior of chiral phase transition we have performed simulations of 2+1 flavor QCD with pion mass ranging from 160 MeV down to 80 MeV
using HISQ action on $N_\tau=6$ lattices. No evidence of first order phase transition in the current quark mass window is found. 
The scaling window of chiral phase transition investigated using the HISQ action shrinks compared to that using the p4fat3 action. However, chiral condensates at all quark masses can be described by the scaling function with a simple regular term. 
This allows a quantitative description of scaling violations for non-zero values of quark masses in the vicinity of chiral phase transition.

\section{Acknowledgements}

The numerical simulations were carried out on clusters of
the USQCD Collaboration in Jefferson Lab and Fermilab, and on BlueGene/L computers at the New York Center for Computational Sciences (NYCCS)
at Brookhaven National Lab. This manuscript has been authored under contract number DE-AC02-98CH10886 with 
the U.S. Department of Energy.

\end{document}